# XQ2P: Efficient XQuery P2P Time Series Processing

BOGDAN BUTNARU[#], BENJAMIN NGUYEN[#], GEORGES GARDARIN[#], LAURENT YEH[#]

[#] *PRiSM Laboratory,*

*45 avenue des Etats-Unis, 78035 Versailles Cedex France*

[1] `firstname.lastname@prism.uvsq.fr`

*Abstract*— **In this demonstration, we propose a model for the management of XML time series (TS), using the new XQuery 1.1 window operator. We argue that centralized computation is slow, and demonstrate *XQ2P*, our prototype of efficient XQuery P2P TS computation in the context of financial analysis of large data sets (>1M values).**

## I. Introduction

Research in time series (TS) [4][6][7] has been very prolific in the last decade. Several domains of applications such as finance, economy, climate evolution, and transport control have been considered.

As proposed in [2], and in the current 1.1 Working Draft [9], XQuery will support a generic construct for TS based on windows, and continuous streams. Window queries can be used in order to carry out algorithmic trading and finding opportunities for arbitrage deals by computing call-put parities, more generally for applying technical analysis to stock chart.

In the prototype presented in this demonstration, we focus on the efficient computation of very large TS of financial data. Our goal is to provide performance comparable to commercial implementations of specific parallel TSDB, such as Vhayu Velocity [10], which is the most popular[1] TS analysis tool in the financial domain. Currently, there is no standard to represent TS and their operations. Most of the time SQL extensions or proprietary languages are used. We advocate the adoption of XQuery 1.1. as standard language to represent TS operations, and present our prototype, XQ2P, which is a >98% XQuery compliant processor implemented with P2P support. We show the feasibility of a TS model appropriate for algorithmic trading using XQuery 1.1, and illustrate our performance gain compared to a generic and centralized XQuery 1.1. implementation (Qizx) and give hints of performance comparison with Vhayu Velocity. XQ2P is implemented on the P2PTester platform [3], in order to provide correct performance measures. The application proposed is real time financial market historical trend computation on multiple series of over 6.000.000 values. In a nutshell, the gain compared to a generic XQuery implementation is **efficiency** through specific P2P optimization, and the gain compared to Vhayu is **integration and generality** by allowing the processing of any XQuery expression involving a general TS (not just financial).

## II. Time series Model

Our basic model is derived from the Roses project [9] adapted and extended for our needs. The model is composed of a vector space of TS equipped with relational-like operations mapping one, two or more TS to one. The model also includes aggregate operators to change the time unit of a series that we do not detail due lack of space. The model also encompasses window-based operations similar to those proposed in the current XQuery 1.1 WD.

### A. Vectors and Vector Space

We define a TS as a potentially infinite vector of values. In the rest of the article, we use *n* to denote the length of the TS. The vector is associated with a calendar giving for each point in time[2] the index of the entry. Time can be of different granularities (e.g., second, day, hour, and week). While in general any kind of XML type, in this article, due to application requirements, values are double precision floats. The calendar starts at a given time which corresponds to the first entry in the associated series; all time units from start to end (the last recorded entry) correspond to an entry. An item is a couple (time, value), i.e., a row in the vector. There exists two possible and distinct null values, the empty (or non-exist) value (denoted "**!**") meaning that there is no value for the given time and the unknown value (denoted "**?**"). TS constitute a linear vector space where addition is denoted + and multiplication by a scalar *. Multiplication and addition of null values are defined as follows, s being a real:

(i) $\begin{cases} !+!=! \\ !+?=? \\ ?+?=? \end{cases}$ (ii) $\begin{cases} s*!=! \\ s*?=? \end{cases}$

TS can be combined linearly in expressions such as $TS_1$ and $TS_2$ are TS (in practice of same calendar and dimensions).

### B. Relational Operators

Logical operators are derived from relational algebra operators specialized for TS. First, the model includes the counterpart of the selection and projection relational

---

[1] According to Vhayu, their software is used by 8 of the top 10 global financial institutions.

[2] There are various possibilities to implement time. Our P2P Java implementation is based on ISO 8601, with arbitrary precision. XML for instance demands at least *ms* precision. Due to lack of space we can not detail further.

operations. Formally, denoting [t,v] the entry t of value v of the processed TS :

$SEL_{pred}(S) = \{[t, v] \mid [t, val] \in S \land v = pred(val)\}$

where pred(val) = val if val satisfies the predicate *pred* and **!** otherwise.

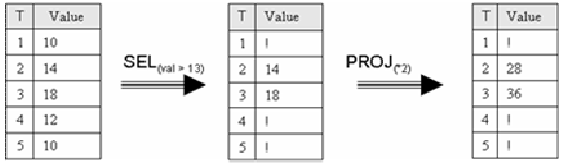

Fig. 1 - Examples of Selection and Projection

We define projection as $PROJ_{fun}(S) = \{[t, m] \mid [t, val] \in S \land m = fun(val)\}$

Examples of Selection and Projection of TS are given in Fig. 1. Note that operators can be composed as with relational algebra to form algebraic expressions. This is true for all operations of our TS algebra. Furthermore, the map function must be defined on null values.

The model also includes some adaptation of the relational outer union and intersection, simply called **union** and **intersection**:

$S_1 \cup S_2 = \{[t, v] \mid [t, v] \in S_1 \lor [t, v] \in S_2\}$
$S_1 \cap S_2 = \{[t, v] \mid [t, v] \in S_1 \land [t, v] \in S_2\}$

Finally, we introduce a k-ary **join** operation for TS based on the same calendar. This operation performs a join on the time attributes of *k* TS using the same calendar, and then applies a mapping function to the tuple of values of the *k* TS : $JOIN_{fun}(S_1, ... S_k) = \{[t, m] \mid [t, val_1] \in S_1 \land ... [t, val_k] \in S_k \land m = fun(val_1, ...val_k)\}$.

JOIN is useful for applications computing derived data from several TS. All these operations find their XQuery counterpart in the *where* clause for selections, and the *let* clause for function calculation.

### C. Window Operator

Most TS applications require a sliding window operator to chop series into consecutive segments and perform an aggregation computation on each segment. Windows are parameterized by their size in number of elements. We use *w* to denote the length of a window.

The sliding window operator computes a series whose $i^{th}$ value is a function of the *w* previous ones, *w* being the window size. Thus, we enrich the space of TS with a generic window-based operator:

$WIN_{fun}(S, w) = \{[t, val] \mid val = fun([t-1, val_1], [t-2, val_2], ...[t-w, val_w])\}$.

Let us recall that [t-i, $val_i$] designates the entry t-i of the TS S of value $val_i$ ; if t-i is negative, $val_i$ is set to $val_0$.

$C_{fun}^{w}$, the cost of computing function *fun* on a window of size *w*.is usually polynomial in *w*. In general $C_{fun}^{TS} \propto C_{fun}^{w} \times n$, where $C_{fun}^{TS}$ is the cost of computing the whole TS. This can be time consuming.

Window operations are implemented in XQuery using the window clause. As illustrated in the following paragraph, functions using these clauses can then be written. For optimization reasons, XQ2P window expressions are run using *specific* code if they respect the TS schema.

### D. Stock Selection and Strategy Evaluation

In finance, *technical analysis* attempts to consider stock prices and volumes as temporal signals and analyse these signals based on indicators, patterns, or events.

Popular window-based operations are the Moving Average (MAVG) and the Relative Strength Index (RSI). MAVG computes the classical moving average series of a series S with a sliding window of size *w*. Let V = $MAVG_w(S)$. The value V[t] of entry t is defined by:

$$V[t] = \sum_{t-w}^{t} \frac{S[t]}{w} = \frac{(w-1)*V[t-1] - S[t-w] + s[t]}{w}$$

A variation is the exponential moving average where value [t-i] is moderated by a weight $(1-alpha)^i$.

There exist many other indicators[8]. Let us stress that the considered operators generate a result TS from the initial TS based on the same calendar. This is useful when developing strategies as introduced below.

Other operators can be computed by combining logical, vectorial, and windowing operators, e.g. Moving Average Convergence/Divergence (MACD). It is one of the simplest indicators used by some investors. A usual formula for the MACD is the difference between a stocks 26-day and 12-day moving averages. Usually, a 9-day moving average of MACD is computed to act as a signal line to buy or sell when crossing 0. The following expression computes the MACD of a series S, then the signal line from the MACD, and finally gives a non empty value supporting a buy decision:

$BUY = SEL_{>0}(MAVG_9(MAVG_{12}(S) - MAVG_{26}(S)))$.

In summary, a financial application requires the ability to run efficient complex expressions with stats operators on long series: a year of quotes at minute resolution is a series of 183.600 entries; a full 15 second precision TS for a year's quotation has a size of 734.300 entries. Queries are functional expressions to compute during a time interval, for example the French stock exchange from 2000/01/01 to 2009/01/01, i.e. 6.615.720 values.

### III. TIME SERIES IMPLEMENTATION

### A. Centralized testing environment

We use a very simple XML schema (Fig. 2) to represent a single value TS. The date element is of type `xs:date` and the value element is an `xs:double`. The functions that we have optimized use this schema as a basis for the TS algebra. In order to process data from other formats we simply transform it so that is conforms to this schema.

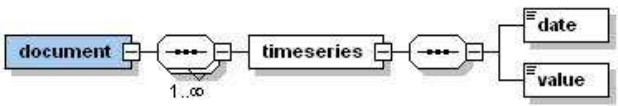

Fig. 2- Timeseries XSD

We have developed XQuery 1.1 functions, using the window clause, to implement the operators previously defined. We also provide specific Java implementation of these functions, to improve performance by bypassing many generic operations of our XQuery implementation. We show as an example the MAVG operator:

```
declare function local:mavg($ts as ts:document,
$i as xs:integer) as ts:document{
<ts:document>{
for sliding window $w in $ts//ts:value start at
$s when fn:true() only end at $e when $e - $s eq
$i  -1
return
<ts:timeseries>
{(data($w/preceding-sibling::ts:date))[$i]}
<ts:value>{avg(data($w))}</ts:value>
</ts:timeseries>
}</ts:document>};
```

Writing a strategy in a declarative way is also very simple. Let us consider MACD (the strategy could also be implemented as a function):

```
let $doc := doc ("lvmh-quotes-ts.xml")/document
let $mavg12 := ts:mavg($doc, 12)
let $mavg26 := ts:mavg($doc, 26)
let $sub := ts:msub($mavg12, $mavg26)
let $macd := ts:mavg($sub, 9)
return
<ts:document>{
for $ts in $macd//ts:value
return
<ts:timeseries>
{$ts/preceding-sibling::ts:date}
<ts:value>{if ($ts > 0) then "buy" else "sell"}
</ts:value>
</ts:timeseries>
}</ts:document>
```

This implementation is very generic and can be enriched simply by programming new classes that compute aggregate functions over a given window of a TS.

### B. Implementing the model in XQuery 1.1

Our centralized tests have been run using Qizx 3.0, because it is to our knowledge one of the few XQuery processors[3] that already supports the 1.1 windowing features. The reasons we chose Qizx over the other processors, are that it is efficient, and that its interface displays the various times spent loading, and processing the query. Nevertheless, Qizx suffers from some limitations, such as the lack of external maths functions, and the fact that it is not a P2P database. On the contrary,

---

[3] Amongst the engines listed on the official XQuery WG page, only MXquery, Zorba, also implement window features. We chose to use Qizx since it was the most efficient.

XQ2P directly implements many missing functionalities for math oriented computing, and of course supports P2P distribution of window computing as shown below.

Our implementation can be easily deployed to a P2P environment since it was developed with distribution in mind using the P2PTester framework [3].

### C. P2P optimization

As TS may be long (e.g., 30 GBs), a peer handling an entire TS might be overloaded, in particular for popular TSs. To avoid this kind of bottleneck, we introduce a method to distribute long TSs into slices on a ring-like addressing space. At loading time, the system distributes TS over the network based on a random hash function. Long TS are split into a sequence of segments. Segments are assigned to peers. Conversely, peers maintain in cache TS segments either imported or calculated. Peers publish the segments they have in cache to other peers by inserting a record in a network DHT (note that we assume this network manages connect, disconnect and replication issues). Every segment has the same length (e.g., 1024 entries for stocks). The last segment is in general incomplete and padded with "**?**" null values. To enable local computation of window-based indicators, we introduce some overlap between segments (e.g., 128 at segment beginning and 128 at segment end for stocks). With such overlap, the local computation of windowing for the core of the series is possible (see Fig. 3).

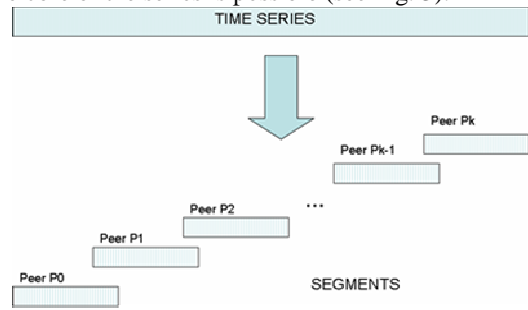

Fig. 3 Distributing TS in segments with overlaps

A derived TS is described by the attributes name, start, and end. Recall that the name is the functional computation tree of the series. For example, CAC40 could be the name of the base TS representing 20.000 days of the French CAC. MAVG(CAC40, 10) is the derived TS obtained by computing the moving average with a window of 10 days. JOIN(MAVG(CAC40,10), SCALE(MOM(CAC40, 5), 100), SUM) is the join of the previous MAVG and the scaling by 100 of the momentum of the CAC40 with a sliding window of 5, using the SUM mapping function. Thus, the name of a derived series with time interval gives all elements to compute the series but also to retrieve parts of the functional tree computing sub-series. This helps us manage a distributed "semantic" cache of TS as explained below.

A problem is that several functional expressions may compute the same TS, for example SCALE(MOM(CAC40, 5),100) gives the same result that

MOM(SCALE (CAC40, 100), 5). This is the classical problem of semantic query rewriting. We define a canonical form of queries to avoid different names for the same derived query.

Every peer shall retrieve relevant segments of a TS efficiently given a name and a time interval. To reach this goal, a DHT-based index is used. The P2P tester provides a Chord implementation; as required by Chord, the keys are hashed to $m$-bit values in an identifier ring of $2^m$ positions. The P2P tester makes possible to map keys to identifiers in a ring-like addressing space using a specific or a standard hashing function. We select a standard hashing function giving approximately the same probability of hit for each ring node (SHA0). The TS name is selected as a key for the DHT and the publishing peers with associated time intervals are recorded in the DHT entry. Thus, publishing a TS in the network is done by the operation *put(key=<name>, content=(<peerId><start><end>)*)*. Keys are unique, but at each publication of the same key, the list is extended. Of course, other approaches are possible.

To avoid re-computing derived TS, we cache on peers the results of expressions for next uses. We assume each peer has a main memory cache with a replacement policy (e.g., FIFO). A peer loading a base TS segment or producing a derived one keeps the series segment in memory cache if possible. For making it available to other peers, it must publish it on the P2P network. This is simply done by performing a *put* in the Chord network as explained above. Notice that all series computed to materialize a functional tree shall be published if kept in cache. Moreover, when a peer removes a TS segment from its cache, it must remove the corresponding entry from the DHT. Thus, all in all, we introduce a cache-based method for TS query processing in a P2P system described in the next subsection.

## IV. Performance evaluation and scenario

Our demonstration scenario covers the calculation of many different strategies, from simple ones shown in this paper to very complex ones, using a dataset of the French stock exchange market (CAC 40) over large datasets (20 years of quotation, approx 6M values per stock). User are invited to test our strategies, or more generally to write *any* valid XQuery 1.1. expression (98% conformance). The P2PTester infrastructure provides detailed performance monitoring of the system (TIME and SPACE), both overall and for each peer used in the demo. While performance is of course optimal with many peers, the demo is illustrated with 4 different physical machines, running 32 peers each.

For a more detailed performance analysis, we refer to [5]. Some Vhayu Velocity performance information can be found in [11]. We simply give as a performance indicator the fact that Vhayu processes **up to 900.000 elements per second** in a 4-octoprocessor server environment with 16GB RAM using 15RPM disks. The following table shows that although we are slower, we achieve similar orders of magnitude with only 128 peers. Our centralized tests have been executed on a Xeon-X5450@3.00GHz with 4GB RAM running Vista-64. Java version is 1.6.0_14 (32 bit) with 1GB heap space on a **1M** length TS. P2P tests have been run using P2P Tester to measure additional routing and network costs.

|  | Qizx 3.0 | XQ2P (on 1 peer) | XQ2P (on 128 peers) |
|---|---|---|---|
| GB Network Load Time | 624ms | 624ms | 624ms + 1140ms routing |
| Bandwidth | 78MB | 78MB | 78MB+1MB routing |
| Disk Load Time | 1986ms | 14370ms | 113ms |
| $WAVG_{100}$ computation | 99 042ms | 1 092ms | 1140ms routing + 128ms computation |
| $MACD_{100}$ computation | 375 000ms (using avg) | 54 819ms | 1140ms routing + 580ms comp. + 624ms netw. |